\documentstyle[11pt,aaspp4]{article}

\def\gs{{_>\atop^{\sim}}}
\def\masomen{{_+\atop^{-}}}

\begin{document}
\title{The High Energy Resolution View of Ionized Absorbers in AGN}
\author{Fabrizio Nicastro}
\affil{Harvard-Smithsonian Center for Astrophysics, 60 Garden Street, 
Cambridge, MA 02138}
\author{Fabrizio Fiore}
\affil{Osservatorio Astronomico di Roma, Via dell'Osservatorio 2, 
Monteporzio Catone (Rm), I-00040 Italy}
\author{Giorgio Matt}
\affil{Universit\'a degli Studi ``Roma-Tre'', Via della Vasca Navale 84, 
I-00146, Roma, Italy}

\begin{abstract}
A number of emission and absorption features are expected to be visible 
in high energy resolution X-ray spectra of type 1 AGN with ionized gas 
along the line of sight (so called ``warm absorbers"). 
Emission strongly depends on the geometrical configuration of the gas, 
while absorption along the line of sight does not. 
Absorption features include photoelectric absorption K and L edges 
along with many strong K$\alpha$,  K$\beta$ and L resonance absorption 
lines from the most abundant elements.
We present detailed simulations of our ``photoelectric + resonant 
absorption'' model with the high energy resolution gratings and 
calorimeters of AXAF, XMM and Constellation-X, and discuss the 
relevant physics which can be addressed with the new generation of 
X-ray spectrometers. 
\end{abstract}

\section{Introduction}
In about 50\% of flat X-ray spectrum, broad optical emission line, 
type 1 AGN, ionized matter  along the line of sight absorbs the 
nuclear X-ray continuum (Reynolds et al., 1997, George et al., 1998). 
Main signatures in the transmitted soft X-ray spectra of these 
AGN are deep and broad K absorption edges, mainly  from highly ionized
Oxygen and Neon, but a number of other weaker and/or narrower absorption 
features are predicted by photoionization and collisional ionization models, 
including Iron L edges, and more than 200 strong K and L resonance 
absorption lines from C, O, Ne, Mg, Si, S and Fe (Nicastro et al., 1998a).  
Emission features are also predicted, with intensities and equivalent widths
highly depending on the geometry. Moreover, these would be the only spectral 
features revealing the presence of this component if the line of sight was not 
obscured by the ionized gas (see also Matt, 1998, this conference). 

However photoionization models predict quite low emission and absorption 
line intensities and equivalent widths, with the latter ranging from $\sim 1$  
to $\sim 10$ eV depending on the geometry and the gas dynamics. 
(Netzer, 1993, 1996, Nicastro, Fiore \& Matt, 1998a, Matt, 1998, this 
conference). 
For this reason moderate resolution soft X-ray spectrometers ($\Delta E = 
0.1$ keV @ 1keV), have so far permitted only the detection and marginal 
separation of the OVII and OVIII absorption K edges at 0.74 and 0.87 keV 
respectively (the strongest predicted absorption features). 
This has allowed estimates of the mean ionization degree and column 
density of the ``warm'' gas (Reynolds et al., 1997). 

\noindent
Many more questions concerning the physical state, the geometry 
and the source of ionization of the gas are still open, and unambiguous 
answers can be found only by the next generation of 
X-ray spectrometers, with greatly improved energy resolution. 
In this work we present our ``Photoelectric+Resonant Absorption'' 
model (Nicastro, Fiore \& Matt, 1998a), and show its diagnostics 
capability in conjuction with high resolution, high collecting area X-ray 
spectra of AGN, as will be available with the next generation of X-ray 
gratings and calorimeters onboard AXAF, XMM and Constellation-X.  

\section{ Models}
Details of our model are described in Nicastro et al. (1998a, 1998b). 
We present here two spectra transmitted from single outflowing 
(v=1000 km s$^{-1}$) clouds of photoionized and collisionally ionized 
gas respectively. The ``turbulence" velocity is $\sigma_v = 1000$ 
km s$^{-1}$ (note that we have not included emission line spectra, since 
these are highly geometry dependent). 
The column density in both cases is fixed at $10^{22}$ cm$^{-2}$ 
(a commonly observed value, Reynolds, 1997) and we adopt an ionization 
parameter (LogU=0.1) and an electron temperature (LogT$_e$=6.5) such that, 
in both cases, the relative ionic abundances of OVII and OVIII are similar. 
We adopt a small covering factor (as seen from the central source) of 
$\Omega/4\pi = 10^{-2}$. This allows us to neglect the contribution 
of the gas emission. 
The 0.1-2.5 keV flux is 2.6$\times 10^{-11}$ erg s$^{-1}$ 
cm$^{-2}$, corresponding to a ROSAT-PSPC count rates of 2.5 ct/s, 
similar to that observed from the brightest Seyfert 1 nuclei (e.g. 
NGC~3516, Mathur et al., 1997). 
Fig. 1 shows these two transmitted spectra. The main resonance absorption 
lines are labeled in both panels.
The different line ratios in the two cases are mainly due 
to the broader ionizing photon distribution (a multi-power law) 
in the photoionization case, compared with the narrow electron 
temperature distribution (a maxwellian) in the collisional case 
(Nicastro et al., 1998b). 
%
%
\begin{figure}
\epsfysize=3.6in
\epsfxsize=3.6in
\vspace{-.8in}
\centerline{\epsfbox{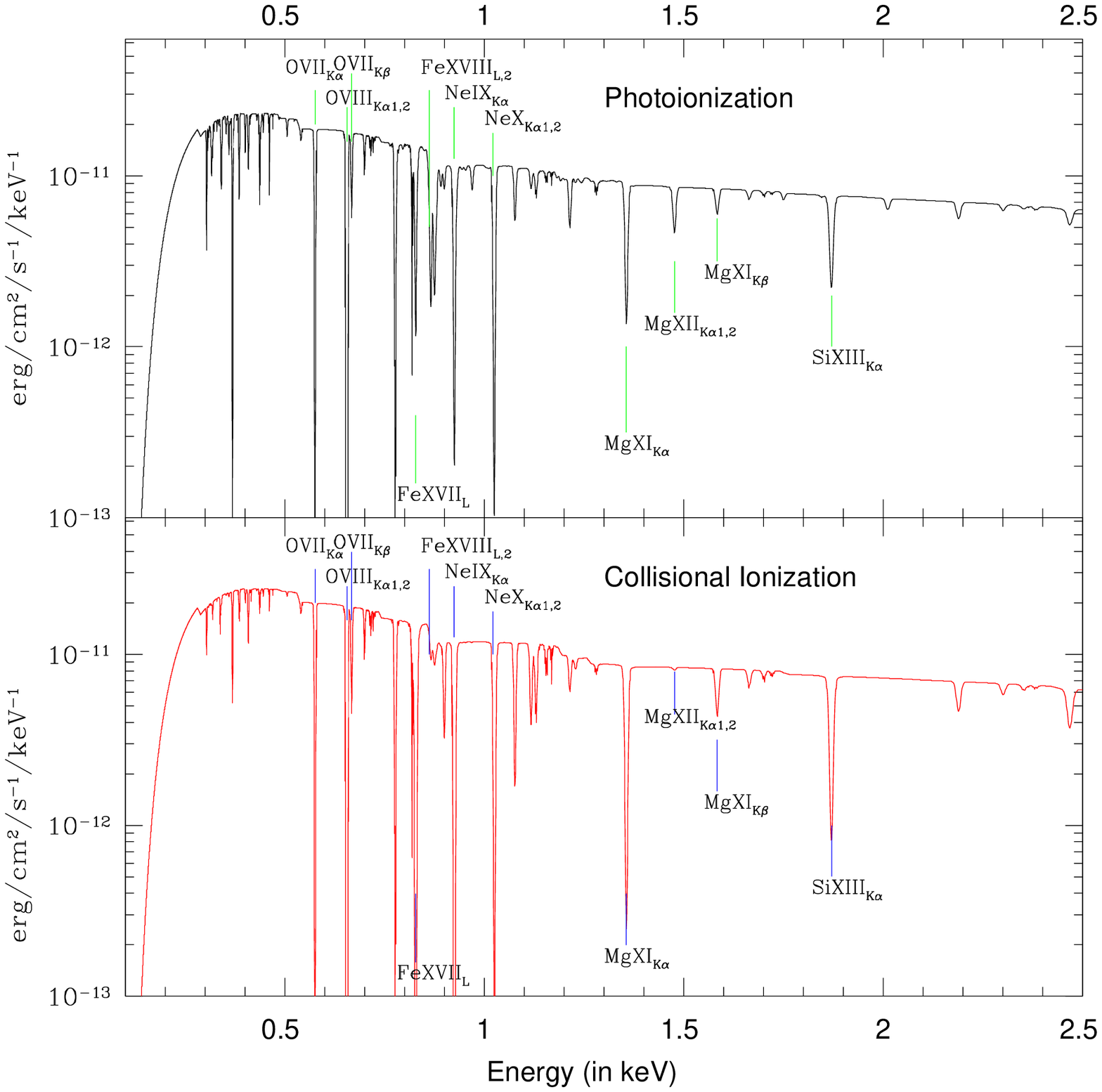}} 
\vspace{-.3in}\caption[h]{Spectra emerging from a cloud of outflowing and 
turbulent ionized gas in photoionization (upper panel) and collisional 
ionization (lower panel) equilibrium. 
Absorption by a column of $3\times 10^{20}$ cm$^{-2}$ neutral gas has been 
included, to account for Galactic absorption.}
\end{figure}

\section{Simulations}
The new generation of X-ray spectrometers will offer the 
opportunity to detect and separate narrow ($\sim 1-2$ eV) absorption 
features. We built a simple ``simulator'' to fold our models with 
the responses of (a) the AXAF-Medium Energy Grating (AXAF-MEG: 
$\Delta E = 1.5$ eV, collecting area = 100 cm$^{2}$ @ 1 keV, 
``AXAF Proposer's Guide'', vs 1.0, 1997); 
(b) the XMM-Reflecting Grating Spectrometer (XMM-RGS 1st order: 
$\Delta E = 3.5$ (1.5 in 2nd order) eV, collecting area $\sim 500$ 
(200 in 2nd order) cm$^2$ @ 1 keV); 
and (c) the baseline Constellation-X calorimeter ($\Delta E = 
3$ eV, collecting area $\sim 10,000$ cm$^2$ @ 1 keV, ``The High Throughput 
X-ray Spectroscopy (HTXS) Mission'', 1997 ). We added statistical 
(and instrumental, when available) noise. 
Fig. 2 shows the 80 ks XMM-RGS simulation (the intermediate case, as far
as the collecting area is concerned).
Spectra of such and even better signal to noise 
will be possible for all the known Seyfert 1 galaxies with ``warm 
absorber''. 
Spectra of comparable quality can be obtained for source with 2-10 flux 
down to $\sim 5 \times 10^{-12}$ erg s$^{-1}$ cm$^{-2}$ ($\sim 600$ targets).
%
%
\begin{figure}
\epsfysize=3.6in
\epsfxsize=3.6in
\vspace{-.8in}
\centerline{\epsfbox{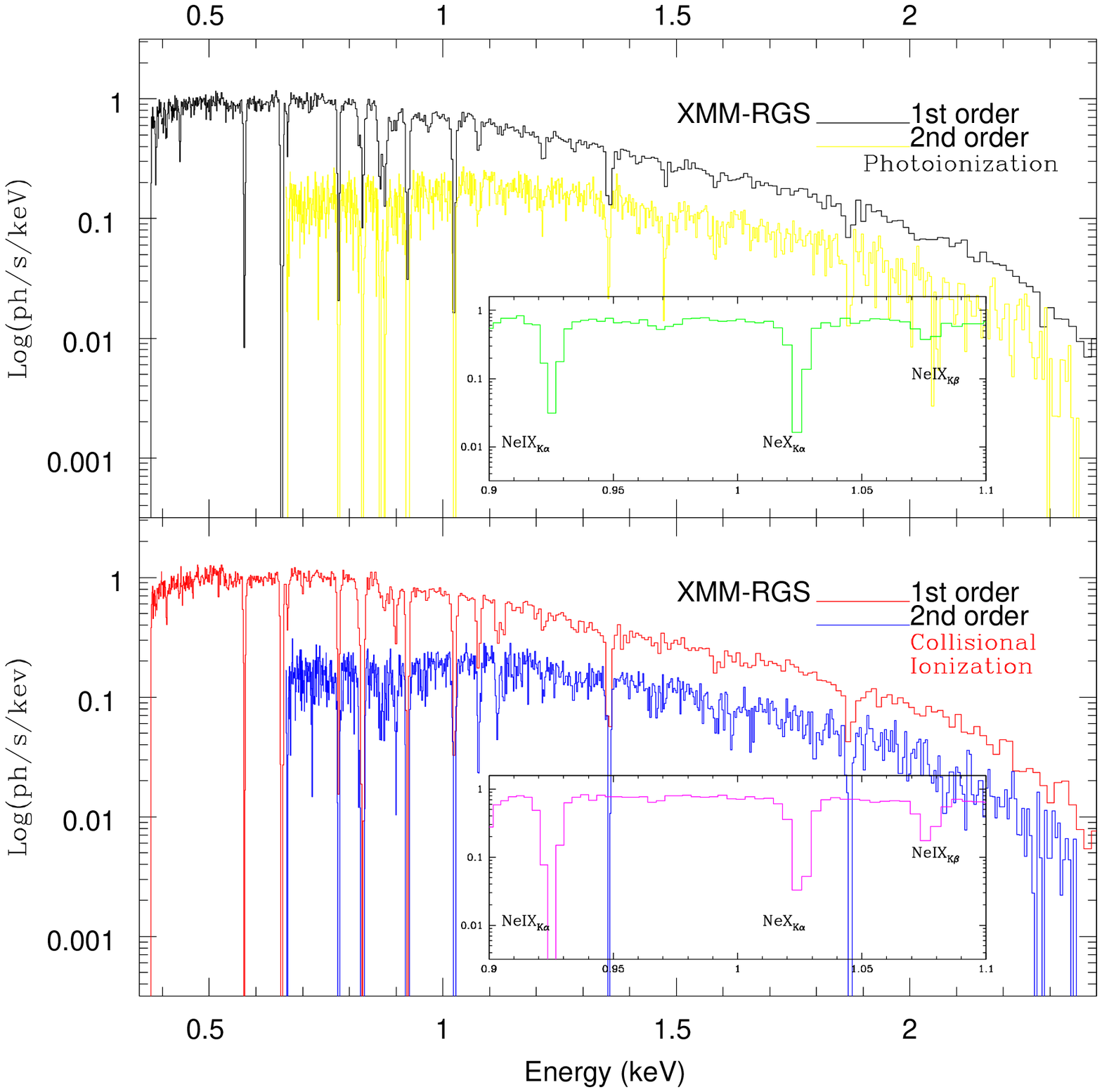}} 
\vspace{-.3in}\caption[h]{80 ks XMM-RGS, 1st and 2nd order order simulations 
of the models in the upper and lower panels of Fig. 1 (upper and lower panels 
respectively). 
Internal windows show the 0.9-1.1 keV portion of the simulated spectra.}
\end{figure}
%
Fig. 3 shows a particular of AXAF-MEG and XMM-RGS (1st and 2nd 
order) simulations of the same model of Fig. 1 (upper panel), for two 
values of the ``turbulence'' velocity $\sigma_v = 200$ and 1000 km 
s$^{-1}$. The bottom panel of this figure shows the same particular 
for a 20 ks Constellation-X simulation and for the case $\sigma_v =
200$ km s$^{-1}$. 
%
%
\begin{figure}
\epsfysize=3.6in
\epsfxsize=3.6in
\vspace{-.8in}
\centerline{\epsfbox{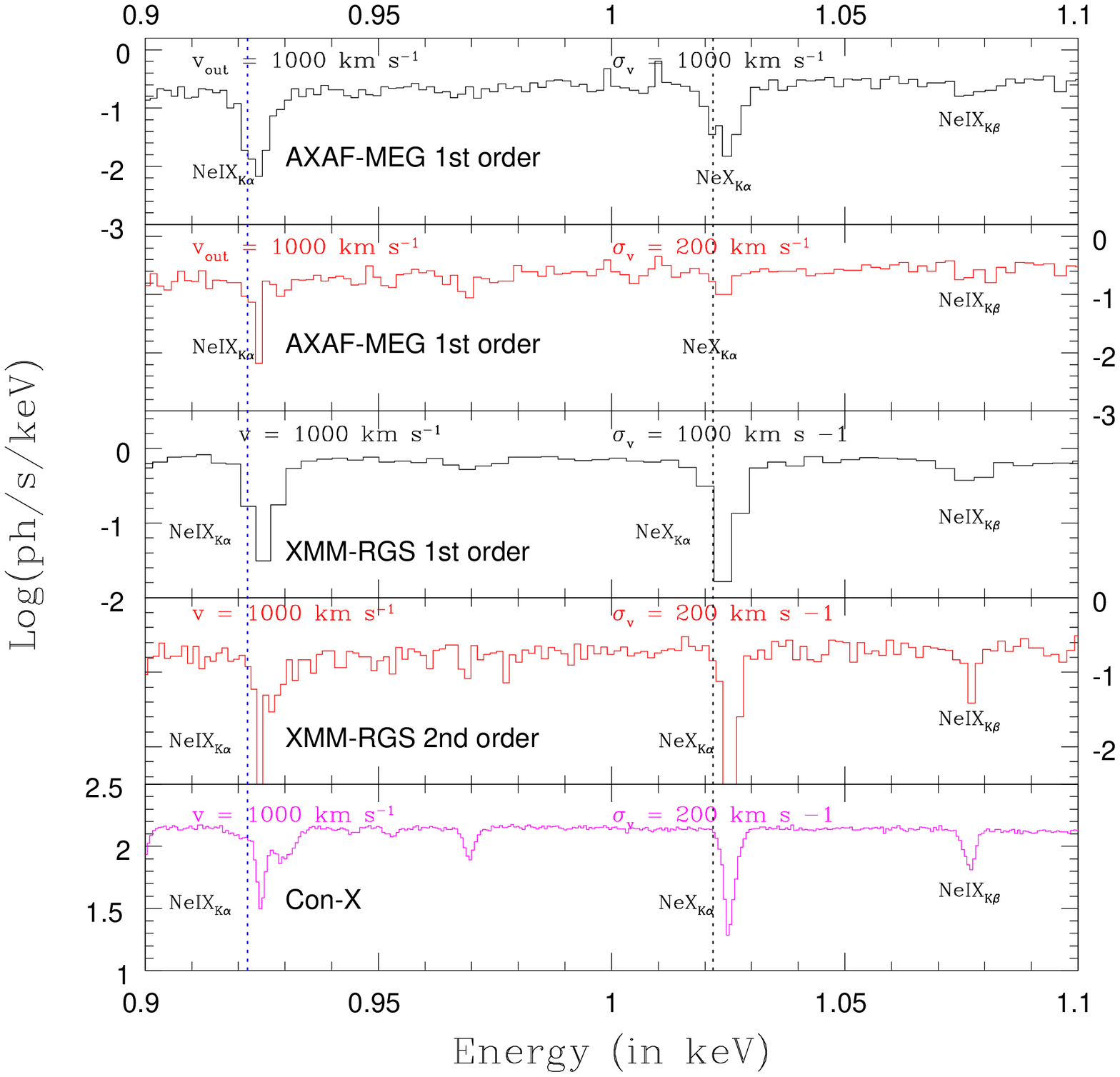}} 
\vspace{-.3in}\caption[h]{0.9-1.1 keV portion of AXAF-MEG  
(first and second panels), XMM-RGS 1st and 2 order (third and fourth 
panel) and Constellation-X (5th panel) simulations of the model in the 
upper panel of Fig. 1. The dashed, blue vertical line in all the panels 
indicates the rest energies of the NeIXK$\alpha$ and the NeXK$\alpha$ 
lines.}
\end{figure}
%

The factor of $\sim 6$ larger collecting area of XMM-RGS, 
compared to AXAF-MEG, will allow detailed spectral variability 
studies of warm absorbers in bright AGN significantly changing their intensity 
state on timescales of $\gs 50$ ks (e.g. NGC~5548). Fig. 4 show this 
capability for two spectra emerging from gas in photoionization 
equilibrium with ionizing continua differing by a factor 2 in intensity. 
%
%
\begin{figure}
\epsfysize=3.6in
\epsfxsize=3.6in
\vspace{-.8in}
\centerline{\epsfbox{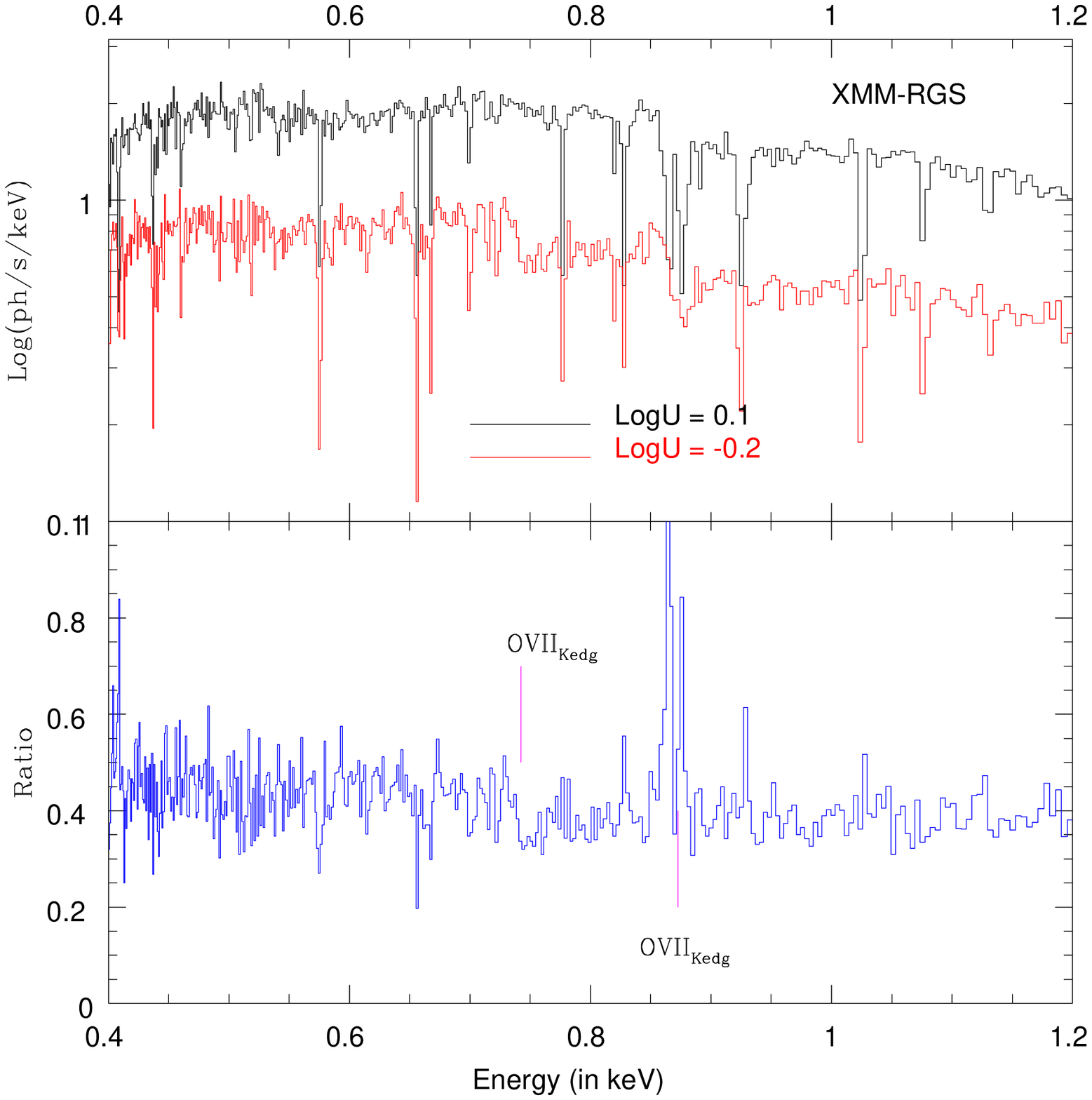}} 
\vspace{-.3in}\caption[h]{80 ks XMM-RGS simulations of two spectra emerging 
from a cloud of gas in photoionization equilibrium with two ionizing continua 
differing by a factor of 2 in intensity (upper panel). 
The lower panel shows the ratio (``Low'' over ``High'' state) between these 
two emerging spectra.}
\end{figure}
%

\section{Diagnostic}

The optical depth $\tau^{X^i}_{\nu^{j \to k}_0}$ of the resonance 
absorption line due to the $j \to k$ transition of the ion $i$ of the element 
$X$, at the core frequency $\nu^{j \to k}_0$ is given by the product between 
the total optical depth and the Voigt profile at the core frequency: 
$\tau^{X^i}_{\nu^{j \to k}_0}=\tau_T \phi_{\nu^{j \to k}_0}$. 
The total optical depth depends linearly on (a) the chemical 
composition A$_X$, (b) the relative ionic abundance $n_X^i$, 
(c) the equivalent hydrogen column density N$_H$, and (d) 
the oscillator strength $f^{X^i}_{j \to k}$: 
\begin{equation}
\tau^{X^i}_{\nu^{j \to k}_0}=\tau_T \phi_{\nu^{j \to k}_0} = 
\left( A_X n_X^i N_H f^{X^i}_{j \to k} \right) \times \phi_{\nu^{j \to k}_0}
\end{equation}

\noindent
A measure of both the optical depth and the relative intensity at 
the center of two observed lines produced by two ions of a same element 
may then allow for an estimate of the relative ionic abundances. 
This in turn allows (a) for unambiguously distinguishing between 
photoionization and collisional ionization (see Table 1, col. II, 
``phot. vs coll''), and (b) for a stringent test of time-evolving 
non equlibrium photoionization model (see Fig. 4, lower panel). 
Furthermore, the measure of the ratio between two lines due to two 
transition of the same ion of the same element, will permit to 
estimate the chemical composition of the gas. 
Finally, the position of the line and its width can give us information 
on the dynamical state of the gas. Accurate measure of blueshift or redshift  
of the absorption lines will allow to estabilish if the gas is 
outflowing or inflowing respectively. 
Furthermore line widths greater than predicted by the only thermal motion 
of the ions will permit to estimate the degree of turbulence of the 
gas along the line of sight (e.g. NeIX, NeX, see Fig. 4). 

\begin{table}[h!]
\caption{Relative ionic abundances, outflowing and turbolence velocity}
\scriptsize
\begin{tabular}{|c|c|ccc|}
\tableline
& Model Predictions & AXAF-MEG & XMM-RGS & Constellation-X \\
\tableline
$(n_{OVIII}/n_{OVII})_{ph}$ & 8.78 & --- & 
$9.1 \masomen 0.8$ & $8.75 \masomen 0.05$ \\
$(n_{OVIII}/n_{OVII})_{col}$ & 7.20 & --- & 
$8.0 \masomen 1.0$ & $7.22 \masomen 0.05$ \\
$(n_{NeX}/n_{NeIX})_{ph}$ & 1.97 & $1.7\masomen0.4$ & 
$2.0 \masomen 0.2$ & $1.98 \masomen 0.02$ \\
$(n_{NeX}/n_{NeIX})_{col}$& 0.50 & $0.6\masomen0.1$ & 
$0.50 \masomen 0.07$ & $0.49\masomen 0.02$ \\
\tableline
\end{tabular}
\tablenotetext{a}{In km s$^{-1}$. Both the outflowing and the turbulence 
velocities are calculated by using the NeIX and NeX K$\alpha$ lines.}
\tablenotetext{b}{Calculated using the 2nd order simulated XMM-RS spectra.}
\end{table}

\noindent
We use the data simulated with the AXAF-MEG, the XMM-RGS (1st and 2nd order) 
and the Constellation-X calorimeter to measure some of the above quantities, 
and present the results in Table 1. The model parameters are those 
described in \S 2, except for the turbulence velocity which is here 
$\sigma_v = 200$ km s$^{-1}$.


\begin{thebibliography}{}

\bibitem[]{} 
Mathur S., Wilkes B.J. \& Aldcroft T., 1997, ApJ, 478, 182.

\bibitem[]{}
George I.M., Turner T.J., Netzer H., Nandra K., Mushotzky R.F., \& Yaqoob T., 
1998, ApJS, 114, 73

\bibitem[]{}
Netzer H., 1993, ApJ, 411-594

\bibitem[]{}
Netzer H., 1996, ApJ, 473, 781

\bibitem[]{}
Nicastro F., Fiore F., Matt G., 1998a, ApJ, submitted.  

\bibitem[]{}
Nicastro F., Fiore F., Perola G.C. \& Elvis M., 1998b, ApJ, in press, 
astro-ph/9808316. 

\bibitem[]{}
Reynolds C.S., 1997, MNRAS 287, 513. 

\end{thebibliography}
\end{document}